\documentclass[conference,10pt]{IEEEtran}
\IEEEoverridecommandlockouts
\usepackage{amsfonts}
\usepackage{amsmath}
\usepackage{amssymb} 
\usepackage{array}
\usepackage{bm} 
\usepackage{bbm}
\usepackage{color} 
\usepackage{cases} 
\usepackage{cite}
\usepackage{dsfont}
\usepackage{graphicx}
\usepackage[bookmarks=false]{hyperref}
\hypersetup{
	colorlinks=true,
	linkcolor=blue,
	filecolor=blue,
	citecolor = blue,      
	urlcolor=cyan,}
\usepackage{hhline}
\usepackage{multirow}
\usepackage{makecell}
\usepackage{stfloats}
\usepackage{textcomp}
\usepackage{url}
\usepackage{units} 
\usepackage{verbatim}

\usepackage[caption=false,font=normalsize,labelfont=sf,textfont=sf]{subfig}
\usepackage[printonlyused]{acronym}
\usepackage[table]{xcolor}

\usepackage{algorithm}
\usepackage{algpseudocode}
\usepackage{tikz} 
\usepackage[utf8]{inputenc}
\usepackage{pgfplots} 
\pgfplotsset{width=10cm,compat=1.9}
\usepackage{pgfplotstable}
\usepackage{xurl} 
\usepackage[colorinlistoftodos,bordercolor=orange,backgroundcolor=orange!20,linecolor=orange,textsize=scriptsize]{todonotes}
\usepackage{yfonts}

\algtext*{EndWhile}
\algtext*{EndIf}
\algtext*{EndFor}
\usetikzlibrary {arrows.meta}
\makeatletter

\makeatletter
\algnewcommand{\LineComment}[1]{\Statex \hskip\ALG@thistlm \(\triangleright\) #1}
\makeatother

\newacro{aoa} [AoA] {angle-of-arrival}
\newacro{aod} [AoD] {angle-of-departure}
\newacro{amp} [AMP] {approximate message passing}
\newacro{arv} [ARV] {array response vector}
\newacro{awgn} [AWGN] {additive white Gaussian noise}
\newacro{cdf} [CDF] {cumulative distribution function}
\newacro{cf} [CF] {cell-free}
\newacro{crb}[CRB]{Cram{\'e}r-Rao bound}
\newacro{crlb}[CRLB]{Cram{\'e}r-Rao lower bound}
\newacro{cs} [CS] {compressed sensing}
\newacro{csi} [CSI] {channel state information}
\newacro{dft}[DFT]{discrete Fourier transform}
\newacro{fa} [FA] {false alarm}
\newacro{fim} [FIM] {Fisher information matrix}
\newacro{ghz} [GHz] {gigahertz}
\newacro{glrt} [GLRT]{generalized likelihood ratio test}
\newacro{iid}[i.i.d.]{independently and identically distributed}
\newacro{isac} [ISAC] {integrated sensing and communications}
\newacro{kld} [KLD] {Kullback–Leibler divergence}
\newacro{lsfc} [LSFC] {large scale fading coefficient}
\newacro{lmmse} [LMMSE] {linear minimum mean-square error}
\newacro{los} [LoS] {line-of-sight}
\newacro{ls} [LS] {least-square}
\newacro{lse} [LSE] {least-square estimator}
\newacro{lb} [LB] {lower bound}
\newacro{map} [MAP] {maximum a posteriori}
\newacro{md} [MD] {missed detection}
\newacro{ml} [ML] {maximum likelihood}
\newacro{mle} [MLE] {maximum likelihood estimation}
\newacro{mmle} [MMLE] {mismatched maximum likelihood estimation}
\newacro{mcrb} [MCRB] {misspecified Cram\'er-Rao bound}
\newacro{mcrlb} [MCRLB] {misspecified Cram\'er-Rao lower bound}
\newacro{mse} [MSE] {mean-square error}
\newacro{mimo} [MIMO] {multiple-input multiple-output}
\newacro{mmwave} [mmWave] {millimeter-wave}
\newacro{mmse} [MMSE] {minimum mean-square error}
\newacro{nlos} [NLoS] {non-line-of-sight}
\newacro{nmse} [NMSE] {normalized mean squared error}
\newacro{np} [NP] {Neyman-Pearson}
\newacro{cpofdm} [CP-OFDM] {cyclic-prefix orthogonal frequency-division multiplexing}
\newacro{pdf}[PDF]{probability density function}
\newacro{rv} [RV] {random variable}
\newacro{rach} [RACH] {random access channel}
\newacro{rss} [RSS] {received signal strength}
\newacro{ru} [RU] {radio unit}
\newacro{ris} [RIS] {reconfigurable intelligent surface}
\newacro{se} [SE] {state evolution}
\newacro{svd} [SVD] {singular value decomposition}
\newacro{snr} [SNR] {signal-to-noise ratio}
\newacro{toa} [ToA] {time-of-arrival}
\newacro{ue} [UE] {user equipment}
\newacro{ula} [ULA] {uniform linear array}
\newacro{upa} [UPA] {uniform planar array}
\newacro{ura} [uRA] {unsourced random access}
\newacro{zc} [ZC] {Zadoff--Chu}

\newcommand{\Tran}{{\sf T}}
\newcommand{\Herm}{{\sf H}}

\newcommand{\SNR}{{\sf SNR}}

\newcommand{\diag}{{\hbox{diag}}}
\newcommand{\blkdiag}{{\hbox{blkdiag}}}

\newcommand{\eqdef}{\stackrel{\Delta}{=}}

\renewcommand{\arg}{{\rm arg}}

\renewcommand{\vec}{{\rm vec}}

\newcommand{\los}{\mathrm{LoS}}

\newcommand{\nlos}{\mathrm{NLoS}}

\newcommand{\spec}{\mathrm{Spec}}
\newcommand{\trun}{\mathrm{trun}}

\newcommand{\av}{{\bf a}}
\newcommand{\bv}{{\bf b}}

\newcommand{\hv}{{\bf h}}

\newcommand{\mv}{{\bf m}}

\newcommand{\pv}{{\bf p}}

\newcommand{\rv}{{\bf r}}
\newcommand{\sv}{{\bf s}}

\newcommand{\vv}{{\bf v}}
\newcommand{\xv}{{\bf x}}

\newcommand{\zerov}{{\bf 0}}

\newcommand{\Am}{{\bf A}}
\newcommand{\Bm}{{\bf B}}
\newcommand{\Cm}{{\bf C}}

\newcommand{\Fm}{{\bf F}}

\newcommand{\Hm}{{\bf H}}

\newcommand{\Jm}{{\bf J}}
\newcommand{\Km}{{\bf K}}

\newcommand{\Rm}{{\bf R}}
\newcommand{\Sm}{{\bf S}}

\newcommand{\Wm}{{\bf W}}

\newcommand{\Xm}{{\bf X}}

\newcommand{\Rrm}{{\rm R}}
\newcommand{\Srm}{{\rm S}}

\newcommand{\Urm}{{\rm U}}

\newcommand{\Acal}{{\cal A}}

\newcommand{\Cc}{{\cal C}}
\newcommand{\Dc}{{\cal D}}

\newcommand{\Gc}{{\cal G}}

\newcommand{\Lc}{{\cal L}}

\newcommand{\Nc}{{\cal N}}

\newcommand{\Pc}{{\cal P}}

\newcommand{\Sc}{{\cal S}}

\newcommand{\muv}{\hbox{\boldmath$\mu$}}

\newcommand{\psiv}{\hbox{\boldmath$\psi$}}

\newcommand{\Sigmam}{\hbox{\boldmath$\Sigma$}}
\newcommand{\Phim}{\hbox{\boldmath$\Phi$}}

\newcommand{\Psim}{\hbox{\boldmath$\Psi$}}

\newcommand{\Cbb}{\mathbb{C}}

\newcommand{\Ibb}{\mathbb{I}}

\newcommand{\Rbb}{\mathbb{R}}

\newcommand{\csf}{{\sf c}}

\newcommand{\xsf}{{\sf x}}

\usepackage{mathabx}

\makeatother

\makeatletter
\def\ps@IEEEtitlepagestyle{%
  \def\@oddfoot{\mycopyrightnotice}%
  \def\@oddhead{\hbox{}\@IEEEheaderstyle\leftmark\hfil\thepage}\relax
  \def\@evenhead{\@IEEEheaderstyle\thepage\hfil\leftmark\hbox{}}\relax
  \def\@evenfoot{}%
}

\def\mycopyrightnotice{%
  \begin{minipage}{\textwidth}
  \centering \scriptsize
    Copyright~\copyright~2026 IEEE. Personal use of this material is permitted. Permission from IEEE must be obtained for all other uses, in any current or future media, including reprinting/republishing this material for advertising or promotional purposes, creating new collective works, for resale or redistribution to servers or lists, or reuse of any copyrighted component of this work in other works.
    
    Accepted for publication in the Proceedings of the IEEE Global Communications Conference (GLOBECOM), 2026.
  \end{minipage}
}
\makeatother

\begin{document}

\bstctlcite{IEEEexample:BSTcontrol}

\title{Random Access and Localization in Cell-Free User-Centric Networks with Multipath Channels\\
\thanks{The work of S. Tarboush was supported by the European Union, through the Horizon Europe Marie Skłodowska-Curie Doctoral Networks Programme “Intelligent sensing and communication as training network for perceptive mobile networks in 6G (ISAC-NEWTON)” under Grant 101169496. The work of E. Gkiouzepi and G. Caire was supported by the BMFTR Germany in the program of “Souverän. Digital. Vernetzt.” Joint Project 6G-RIC (Project IDs 16KISK030). All authors contributed equally to this work.
}
}

\author{
    \IEEEauthorblockN{
        Simon Tarboush,
        Eleni Gkiouzepi,	
		and Giuseppe Caire
        }
	   \IEEEauthorblockA{
        \small Communications and Information Theory Chair, Faculty of Electrical Engineering and Computer Science,\\ Technische Universit{\"a}t Berlin, 10587 Berlin, Germany.
        Email: \{simon.tarboush,gkiouzepi,caire\}@tu-berlin.de).
        }
}

\maketitle

\begin{abstract}
In a wireless network, the initial/random access mechanism (RACH) allows idle/new users to join the network and (possibly) request allocated transmission resources for subsequent traffic. 
Building on our own previous work, for cell-free user-centric networks, we consider location-dependent random access codebooks such that users in a certain geographic area (location) make use of the corresponding set of random access preambles (codewords). 
We expand our previous work in two ways: 1) we consider multipath channels with line-of-sight (LoS) propagation within a given radius. 2) We consider two different approaches. The first makes use of Zadoff–Chu (ZC) sequences and GLRT detection to cope with the unknown delay, and it is conceptually similar to the 3GPP 2-step RACH specification (here extended to the cell-free case). The second builds on our previous work on multisource approximate message passing (AMP). For both schemes, we also consider a novel {\em near Maximum-Likelihood} approach for localization of the random access users directly from the detected RACH preambles, implicitly using angle of arrival and time difference of arrival information embedded into the LoS components. Simulation results show that the AMP approach achieves generally better performance for random access user detection, while both approaches have similar localization capability with a slight superiority for the frequency-domain scheme.
\end{abstract}

\section{Introduction}

Random access is a key functionality in any wireless network, allowing idle and new users to join the network. The \ac{rach} mechanism makes use of a set of codewords (preambles) that the users can transmit in order to signal their presence to the network. Such codewords are not uniquely assigned to the users (in fact, there are several billions of wireless devices in the world). Instead, all users make use of a common codebook, and pick one of the possible preambles with uniform probability. In this way, preamble collision is mitigated. A random access channel with virtually infinite users, where only a finite number of users transmits at any time slot making use of a common codebook, is referred to as \ac{ura}~\cite{Polyanskiy2017Perspective,liva2024unsourced,Ozates2025Unsourced} in the information theoretic literature.
As a matter of fact, virtually any standardized \ac{rach} mechanism (e.g., the two-step \ac{rach} 5G NR~\cite{ETSI138}) belongs to this class. The network scans the \ac{rach} slots and detects the active preambles, which act as ``token'' signaling the presence of some user that wants to transmit data and possibly request an allocated transmission resource for subsequent connected traffic. In \ac{cf} user-centric networks (see \cite{Ngo2024Ultradense,Demir2021Foundations} and references therein), any user joining the network must be assigned a user-centric cluster of \acp{ru} serving such user. Hence, it is highly desirable that random access users can also be accurately localized from their transmission in the 
\ac{rach} slot. 

Building upon the idea of a location-based partition of the coverage area into disjoint zones, originally proposed by our research group \cite{Cakmak2025Joint,Gkiouzepi2024Joint}, such a partition aims to resolve the ambiguity of the unknown association between the transmitted codewords and the \acp{lsfc} of the propagation environment between the active users and the \acp{ru}. In this work, we first take a step further in the channel model used, where we adopt a frequency-selective spatially correlated fading model and ensure spatial consistency of the channels generated among different users, since for any given area served by multiple \acp{ru} it is expected that users will share a common scattering environment and will have a partially overlapping structure. Such channel modeling is often overlooked in the current \ac{cf} literature~\cite{Meng2025Network,He2025RSS} even in our previous work~\cite{Cakmak2025Joint,Gkiouzepi2024Joint,Gkiouzepi2025Joint}. Within this framework, we consider two receiver pipelines. The first is a time-domain receiver based on \ac{zc} sequences and a \ac{glrt}, which is closely aligned with the current \ac{rach} design. The second is a frequency-domain receiver based on multisource \ac{amp}~\cite{Cakmak2025Joint}, which jointly exploits the observations collected across all \acp{ru} in the network, where both schemes have used the idea of a geographically partitioned access codebook.

The main contribution is a unified comparison between these two random-access strategies under the same ``channel load'' (i.e., average number of active random access users per unit area) in a realistic frequency-selective spatially-consistent propagation. The system performance is evaluated in terms of the probability of detection of the random access users and position estimation. Numerical simulation confirms that the AMP-based approach provides superior performance with respect to the ``legacy'' \ac{zc} scheme in terms of detection and provides better localization accuracy. 

{\bf Notation:} $\av$ and $\underline{\av}$ denote column and row vectors. The multivariate circularly symmetric complex Gaussian distribution with mean $\mv$ and covariance $\Cm$ is denoted by $\Cc\Nc(\mv,\Cm)$, while its density function is denoted by $\textswab{g}(\cdot; \mv, \Cm)$.

\section{System and Channel Models}
\label{sec:sys_model}

\subsection{System Model}

\begin{figure}[ht]
	\centering 	\includegraphics[width=0.47\textwidth]{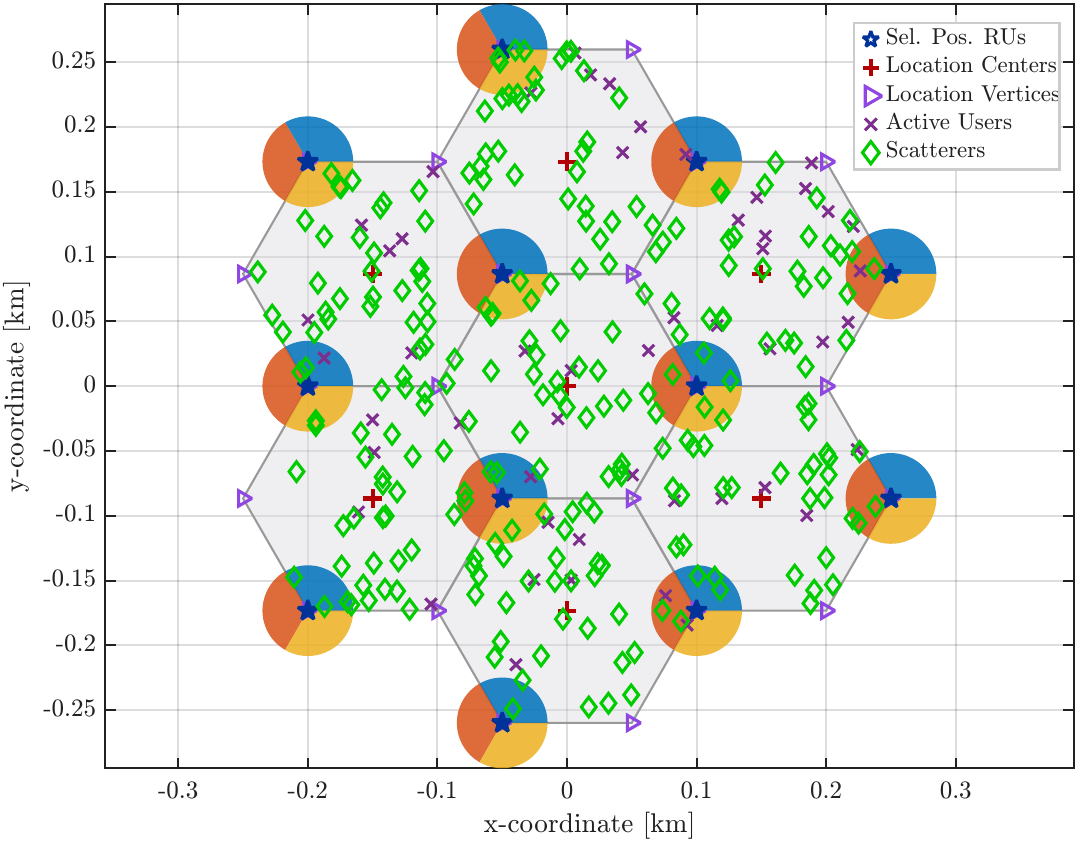}
	\caption{Example of the considered network showing the location of \acp{ru} and a random realization of the active users and scatterers.}
    \label{fig:network_topology}	
\end{figure}

We consider a \ac{cf} user-centric wireless network with $B$ \acp{ru} at fixed, known locations $\pv_b \in \Rbb^2 : b \in [B]$, each equipped with $M$ elements \ac{ula}, serving a population of single-antenna random access users with unknown (to the network) positions $\pv_k \in \Dc \subset \Rbb^2$, where $\Dc$ is the network coverage area.
In the same region $\Dc$, a total of $S$ scatterers are placed at positions $\pv_s$. 
For the purpose of the \ac{rach}, the coverage area $\Dc$ is partitioned into $U$ disjoint zones $\Dc_u: u \in [U]$, referred to as ``\textit{locations}''.
The layout considered in our numerical results is shown in Fig.~\ref{fig:network_topology}, where 12 sites are placed on a hexagonal lattice. Each site contains 3 \acp{ru}, each covering a 120$^\circ$ angle (indicated by the three colored sectors) for a total of $B = 36$ \acp{ru}. The coverage area (in gray) is partitioned into $U = 7$ locations $\Dc_u$ corresponding to the hexagons. Fig.~\ref{fig:network_topology} also shows a realization of the randomly placed active users and scatterers.

The general \ac{rach} scheme works as follows \cite{Cakmak2025Joint,Gkiouzepi2024Joint,Gkiouzepi2025Joint}: each location $\Dc_u$ is assigned a distinct uRA codebook formed by $N$ codewords (\ac{rach} preambles) of length $L$ ``chips'' for a total duration of $\approx L T_s$ where $T_s = 1/W$ and $W$ is the system bandwidth, arranged for convenience as a matrix $\Sm_u \in \Cbb^{L \times N/U}$, where  $N$ is the total number of codewords in the system and $\Sm = [\Sm_1, \ldots, \Sm_U]$ is the total codebook matrix. Users can determine their location from a trigger beacons transmitted periodically by all the RUs simultaneously to indicate the beginning of the \ac{rach} slot (e.g., this can be done by mapping the vector of received beacon powers from the different RUs into the location index as described in \cite{Gkiouzepi2025Joint,testi2026weighted}). When a user in location $\Dc_u$ wishes to access the \ac{rach}, it chooses a codeword in codebook $\Sm_u$ and sends it at given power (fixed by the scheme) in the \ac{rach} slot. 
The system obtains the received signal in correspondence of the \ac{rach} slot from all RUs and must identify the ``active'' codewords (i.e., the codewords that have been effectively transmitted) and the position of the active users with more accurate positioning with respect to the (typically very coarse) locations. The particular format of the uRA codebooks and the corresponding detection and localization algorithms is the main contribution of this paper and will be treated later. 

The users may choose their codeword depending on some information bits \cite{fengler2022pilot} or simply at random \cite{ETSI138}. This aspect is irrelevant for the purpose of this paper and shall not be discussed. 
Furthermore, some codewords may be chosen by more than one user. Such {\em preamble collisions} impact the overall \ac{rach} performance \cite{liva2024unsourced}, depending on the particular collision resolution scheme. Since this source of impairment is quite different in nature from the detection error events, in our results, we assume no collisions. It is generally true that large codebooks can mitigate preamble collision and should be preferred. Schemes that estimate the codeword multiplicity, i.e., not only whether a codeword was or was not transmitted, but also how many users have transmitted the same codeword, have been studied in \cite{krishnan2025achievability}.

\subsection{Channel Model}

To illustrate the channel model, we indicate with the index $k$ a generic active user placed at $\pv_k \in \Dc_u$. 
The continuous-time channel vector $\hv_{b,k}(\tau) \in \Cbb^M$ from user $k$ and \ac{ru} $b$ at $\pv_b$ is given by $\hv_{b,k}(\tau) = \hv^\los_{b,k}(\tau) + \hv^\spec_{b,k}(\tau)$. The \ac{los} component takes on the form
\begin{equation}
    \label{eq:los_comp}
    \hv^\los_{b,k}(\tau) = \Ibb^\los_{b,k} \sqrt{\beta^\los_{b,k}} {\rm e}^{j \varphi_{b,k}} \av(\theta_{b,k})g(\tau-\tau^\los_{b,k}),
\end{equation}
where $\beta^\los_{b,k}$ is the distance-dependent \ac{lsfc} of the \ac{los} path, $\varphi_{b,k} \sim {\rm Unif}[0,2\pi)$ is a uniform random phase to model the fact that any small change, even of the order of half a wavelength, in the user location may result in a significant phase-shift of the \ac{los} component~\cite{Ozdogan2019Performance,Wang2020Uplink}, $\tau^\los_{b,k} = d_{b,k}/\csf$ is the propagation delay, where $d$ indicate distance and $\csf$ is the speed of light, $g(\cdot)$ is the overall Nyquist pulse shape (convolution of the chip transmit pulse and its corresponding matched filter), and $\av(\theta_{b,k}) \in \mathbb{C}^M$ denotes the \ac{ru} far-field array vector as a function of the local \ac{aoa} $\theta_{b,k}$.\footnote{In this work we consider $\lambda/2$ spaced ULAs at the RUs.}  The indicator function $\Ibb^\los_{b,k} \in \{0,1\}$ in \eqref{eq:los_comp} controls the presence of the \ac{los} path. Specifically, a \ac{los} component exists if and only if i) the $d_{b,k} \leq d^\trun_{\Rrm,\Urm}$ (i.e., user $k$ is within a certain truncation radius from RU $b$) and $\theta_{b,k} \in \Acal_b$ (the \ac{aoa} falls within the \ac{ru}'s angular sector). 

The specular \ac{nlos} component represents a distinct reflection path through a large surface (compared to wavelength) of the randomly distributed scatterers.\footnote{Here we consider only single-bounce reflections.}
A \ac{nlos} path from active user $k$ to RU $b$ through scatterer $s$ exits if 
\begin{equation}
d_{s,k} \le d^\trun_{\Srm,\Urm} \;\land\; d_{b,s} \le d^\trun_{\Rrm,\Srm} \;\land\; \theta_{b,s} \in \Acal_b
\label{pathexistence}
\end{equation}
where $\theta_{b,s}$ is the \ac{aoa} of scatterer $s$ with respect to RU $b$, and where $d^\trun_{\Srm,\Urm}$ and $d^\trun_{\Rrm,\Srm}$ are system parameters. Let $\Sc_{b,k}^\spec$ denote the set of scatterers satisfying \eqref{pathexistence}. Then, the \ac{nlos} channel between user $k$ and \ac{ru} $b$ is given by 
\begin{equation}
\label{eq:spacular_comp}
    \hv^\spec_{b,k}(\tau) =\sum_{s \in \Sc_{b,k}^\spec} \sqrt{\beta_{b,k,s}^\spec} \rho_{b,k,s} \av(\theta_{b,s}) g(\tau-\tau^\nlos_{b,k,s}),
\end{equation}
where $\beta_{b,k,s}^\spec$ is the distance-dependent pathloss, $\rho_{b,k,s} \sim\Cc\Nc(0,1)$ are mutually independent for different indices, $\tau^\nlos_{b,k,s} = (d_{k,s} + d_{s, b})/\csf$ is the path delay, and $\av(\theta_{b,s})$ is the array vector. Notice that, for both \ac{los} and \ac{nlos} components, delays, \ac{lsfc}, and \ac{aoa} are functions of the position of the user (and scatterers) w.r.t. the \ac{ru}. 
Following this approach, the adopted channel modeling ensures spatial consistency across the whole network. 

The discrete-time channel is obtained by sampling $\hv_{b,k}(\tau)$ at integer multiples of the chip interval $T_s$. Hence, for each $k$ and $b$ we have $M$ discrete-time impulse responses of finite duration. 
In particular, we consider a rectangular chip pulse such that $g(\tau)$ is a triangular pulse of duration $2 T_s$, defined 
\begin{equation}
\label{eq:g_sampled}
  g\!\left(\!\frac{\ell - \tau_p W}{W}\!\right)
  \!=\!
  \begin{cases}
    \ell - \tau_p W,       & \ell \in [\tau_p W,\;\tau_p W+1), \\
    2 + \tau_p W - \ell,   & \ell \in [\tau_p W+1,\;\tau_p W+2], \\
    0,                     & \text{otherwise,}
  \end{cases}
\end{equation}
where each path contributes at most \emph{two} non-zero taps.
Hence, the discrete-time channel length in samples satisfies $D  \leq {\rm CP}$, where ${\rm CP} = \lfloor \Delta \tau/T_s  + 2 \rfloor$ and $\Delta\tau$ is the delay spread. Then, for a generic path $p$, by ordering delays as $\tau_0^\los \leq \tau_1^\nlos \leq \cdots \leq \tau_P^\nlos$, defining the integer delay as $\ell_p = \lceil \tau_p W \rceil$ and letting $\ell_0 = \ell^\mathrm{min}_{b,k}$ be the smallest one; the channel has non-zero taps only for $\ell \in \{\ell_0,\ldots,\ell_0+D\}$. The fractional delay is $\mu_{p} = \ell_{p} - \tau_{p} W$, such that $\mu_{p} \in [0, 1)$. The $L\times M$ time-space channel matrix, after zero padding to length $L$, can be written
\begin{equation}
  \Hm^\circ_{b,k} = \Jm^{\ell_0}\Hm_{b,k},
  \label{eq:H_factored}
\end{equation}
where the first $D+1$ rows contains the $M$ antenna taps at that lag, $\Jm$ is the $L\times L$ cyclic downshift matrix, and $\Hm_{b,k} \in \Cbb^{L\times M}$ collects all non-zero discrete-time channel taps. We define the $L\times L$ \emph{ downshift}
matrix $\Jm$ as
\begin{equation}
  \Jm
  =
  \begin{bmatrix}
    \mathbf{0}_{L-1}^\mathsf{T} & 1 \\
    \mathbf{I}_{L-1}            & \mathbf{0}_{L-1}
  \end{bmatrix}.
  \label{eq:downshift}
\end{equation}

For any vector $\sv = (s[0], s[1], \ldots, s[L-1])^\Tran$, $\Jm\sv$ cyclically shifts it down by one position and $\Jm^\ell\sv$ shifts by $\ell$ positions. The circulant matrix built on $\sv$ is
\begin{equation}
  \operatorname{Circ}(\sv) =  \bigl[\sv,\,\Jm\sv,\,\Jm^2\sv,\ldots,\Jm^{L-1}\sv\bigr] = \sum_{\ell=0}^{L-1} s[\ell]\,\Jm^\ell.
  \label{eq:circ_def}
\end{equation}
The unitary \ac{dft} matrix $\Fm \in \Cbb^{L \times L}$ has entries $[\Fm]_{c,r}\eqdef\frac{1}{\sqrt L}\omega^{c r}$ with $\omega={\rm e}^{-j2\pi/L}$ and $c, r = 0, \ldots, L-1$. Then, we have the following properties
\begin{align}
    \operatorname{Circ}(\sv)&= \Fm^\Herm\diag(\check\sv)\Fm, \,\text{with} \,\, \check\sv\eqdef\sqrt{L}\Fm\sv,\label{eq:cir_mat_dft}\\
    \Fm\Jm^\ell&=\Phim^\ell\Fm, \label{eq:F_Phi_J}
\end{align}
where $\Phim \eqdef \mathrm{diag}(1,\omega,\ldots,\omega^{L-1})$ denotes the linear-phase shift matrix.

\subsection{Received Signal Models}
\label{subsec:rx_models}

In the collision-free regime, each active index $k$ corresponds to a unique active user. 
\subsubsection*{Time-domain ZC signaling with CP}
Let $\mathbf{s}_k\in\mathbb{C}^{L}$ denote the time-domain codeword transmitted by active user $k$  (in particular, a ZC sequence) and let $\mathbf{H}_{b,k}\in\mathbb{C}^{L\times M}$ denote the matrix containing by columns the $M$ discrete-time impulse responses of the channel from user $k$ to RU $b$, one column for each RU antenna, zero-padded to length $L$. The signal bandwidth is assumed to be larger than the channel coherence bandwidth, resulting in a frequency-selective channel model. Hence, to ensure circular convolution in the presence of a frequency-selective channel, this scheme uses a cyclic prefix of length ${\rm CP}$ so that the receiver may discard the first ${\rm CP}$ samples and obtain a block of length-$L$ affected by circular convolution with the delay-domain channel. Hence, the received length-$L$ \ac{rach} block at RU $b$ after CP removal is
\begin{equation}
\mathbf{Y}_b = \sum_{k=1}^{N} a_k\,\operatorname{Circ}(\mathbf{s}_k)\,\Jm^{\ell_{b,k}} \mathbf{H}_{b,k} +\mathbf{W}_b,
\label{eq:td_rx_model}
\end{equation}
where $a_k\in\{0,1\}$ is the activity indicator, $\operatorname{Circ}(\mathbf{s}_k)$ is the cyclic convolution matrix following \eqref{eq:circ_def}, $\ell_{b,k}$ is the quantized (discrete-time) delay of the shortest path between user $k$ and RU $b$, and $\mathbf{W}_b\in\mathbb{C}^{L\times M}$ contains i.i.d. $\mathcal{CN}(0,\sigma_w^2)$ entries.

\subsubsection*{Frequency-domain CP-OFDM signaling}

For each subcarrier $\xi$, time-frequency domain pilot symbol $q$, and antenna $m$, the received frequency-domain
sample at RU $b$ is
\begin{equation}
y_{b,m}^{(q)}[\xi] = \sum_{k=1}^{N} a_k\, s_k^{(q)}[\xi]\,\tilde h_{b,m,k}[\xi] + w_{b,m}^{(q)}[\xi],
\label{eq:fd_scalar_model}
\end{equation}
where $s_k^{(q)}[\xi]$ is the CP-OFDM pilot symbol transmitted by index $k$ at pilot time $q$ and subcarrier $\xi$, $\tilde h_{b,m,k}[\xi]$ is the frequency domain channel components obtained by the discrete Fourier transform of the sampled version of $\hv_{b,k}(\tau)$, and $w_{b,m}^{(q)}[\xi]\sim\mathcal{CN}(0,\sigma_w^2)$. The frequency-domain received signal is obtained by taking the \ac{dft} of \eqref{eq:td_rx_model}, recall \eqref{eq:H_factored} and using the properties \eqref{eq:cir_mat_dft}, \eqref{eq:F_Phi_J} applied to a channel zero-padded to length $L_f$. Because the \ac{dft} is unitary, $\tilde{\Wm}_b$ retains the same $\mathcal{CN}(0,\sigma_w^2)$ distribution. In the CP-OFDM transmission format, $Q$ consecutive blocks of $L_f$~chips (plus CP) replace the single long block of time-domain signal, and the per-subcarrier scalar observation follows from restricting the above to $L_f$ subcarriers, such that the two schemes have the same overall transmission length in chips.
Stacking the $Q$ pilot observations across the $M$ antennas of RU $b$ yields $\tilde{\mathbf{Y}}_b[\xi]\in\mathbb{C}^{Q\times M}$ defined 
\begin{equation}
\tilde{\mathbf{Y}}_b[\xi] = \check{\mathbf{S}}[\xi]\mathbf{X}_b[\xi]+\tilde{\mathbf{W}}_b[\xi],
\label{eq:fd_matrix_model}
\end{equation}
where $\check{\mathbf{S}}[\xi]\in\mathbb{C}^{Q\times N}$ contains the pilot signatures at subcarrier $\xi$, and the $k$th row of $\mathbf{X}_b[\xi]\in\mathbb{C}^{N\times M}$ is $a_k\, {e}^{-j\frac{2\pi \xi}{L_f}\ell_{b,k}} \tilde{\mathbf{h}}_{b,k}^{\mathsf T}[\xi]$. Define $\underline{\breve{\mathbf{h}}}_{b,k}[\xi] \eqdef {e}^{-j\frac{2\pi \xi}{L_f}\ell_{b,k}} \tilde{\mathbf{h}}_{b,k}^{\mathsf T}[\xi]$, denote $\Pc_{b}$ the set of multipaths satisfying \eqref{pathexistence}, and by dropping the index $k$ for notation simplicity, the channel vector $\underline{\breve{\hv}}_{b}[\xi] \in \Cbb^{1 \times M}$ is 
\begin{equation}
\begin{aligned}
    \underline{\breve\hv}_{b}[\xi] &= \Ibb^\los_b e^{j\varphi_{b}} \!\sqrt{\beta_{b}^\los} v_{b,0}[\xi]\,\underline\av(\theta_{b}^\los)+\\&\sum_{p\in\Pc_b}\!\rho_{b,p}\sqrt{\beta_{b,p}^\spec} v_{b,p}[\xi] \underline\av(\theta_{b,p}).
\label{eq:fd_channel_decomp}
\end{aligned}
\end{equation}

Concatenating \eqref{eq:fd_channel_decomp} across $\xi=0,\ldots,L_f-1$ defines the $L_f\times M$ frequency-space channel matrix as
\begin{equation}
\begin{aligned}
    \breve\Hm_{b} &= \Ibb^\los_b \sqrt{\beta_{b}^\los}\,e^{j\varphi_{b}}\,\Phim_{b,L_f}^{\ell_0}\bv_b(\mu_0)\underline\av(\theta_{b}^\los) + \\& \sum_{p\in\Pc_b}\rho_{b,p}\sqrt{\beta_{b,p}^\spec}\,\Phim_{b,L_f}^{\ell_p}\bv_b(\mu_p)\underline\av(\theta_{b,p}),
\label{eq:breveH}
\end{aligned}
\end{equation}
where $\vv_{b}(\ell,\mu) \eqdef \Phim_{b,L_f}^{\ell} \bv_{b}(\mu)$ contains two parts, a diagonal matrix $\Phim_{b,L_f}^{\ell}\in \Cbb^{L_f \times L_f}$ containing the linear phase shifts across subcarriers due to the integer propagation delay and $\bv_{b}(\mu) \in \Cbb^{L_f}$ is the frequency-domain representation of the fractional delays due to pulse shaping and sampling, with 
\begin{equation}
\begin{aligned}
v_{b,p}[\xi] &= e^{-j\frac{2\pi}{L_f}\ell_{b,p}\xi}\,b_{\mu_{b,p}}[\xi], \\
\text{with} \; b_{\mu_p}[\xi] &= \mu_p+(1-\mu_p)e^{-j\frac{2\pi}{L_f}\xi}.
\label{eq:frac_delay_freq}
\end{aligned}
\end{equation}

\section{Active codewords detection and user positioning}
\label{sec:joint_det_est}

In this section, we describe the active codeword detection for the time-domain and frequency-domain schemes and the proposed localization approaches.

\subsection{Time-domain scheme}
\label{subsec:td_glrt}

The receiver exploits the quasi-perfect cyclic autocorrelation and low cyclic cross-correlation of ZC sequences:
\begin{align}
\operatorname{Circ}(\mathbf{s}_k)^{H}\operatorname{Circ}(\mathbf{s}_k)
&\approx
E_s\,\mathbf{I}_L,
\nonumber\\
\operatorname{Circ}(\mathbf{s}_k)^{H}\operatorname{Circ}(\mathbf{s}_{k'})
&\approx
\mathbf{0},
\quad k'\neq k,
\label{eq:zc_corr_td}
\end{align}
where $E_s=  \|\mathbf{s}_k\|^2$.
Applying the matched filter associated with $\mathbf{s}_k$ gives
\begin{equation}
\mathbf{Y}_{b,k}^{\mathrm{mf}}
= 
\operatorname{Circ}(\mathbf{s}_k)^{H}\mathbf{Y}_b 
\approx
a_k E_s\,\Jm^{\ell_{b,k}} \mathbf{H}_{b,k}
+
\mathbf{W}'_{b,k},
\label{eq:mf_out_td}
\end{equation}
where $\mathbf{W}'_{b,k}\in\mathbb{C}^{L\times M}$ collects noise plus residual multiuser interference caused by the non-ideal cyclic cross-correlations.
The detector does not know (yet) the exact active user position, but only the association of sequences to locations $\Dc_u$ due to the uRA codebook partitioning. 
For a given $\sv_k \in \Sm_u$, in order to detect whether $a_k = 0$ (sequence $\sv_k$ not transmitted) or $a_k = 1$ (sequence $\sv_k$ transmitted), the detector uses the observations at a subset of RUs $\Sc_k$ chosen in the surroundings of location $\Dc_u$. In particular, for the layout of Fig.~\ref{fig:network_topology} for each $\Dc_u$ (hexagon) we choose the three RUs at its vertices whose sector faces the hexagon.  This leads to the generally mismatched (due to imperfect and incomplete a priori information) composite hypothesis testing 
\begin{subequations}  \label{composite-hyppo}
\begin{align}
\mathcal{H}_{0,k}:&\quad
\mathbf{Y}_{b,k}=\mathbf{W}'_{b,k},   \;\;\; \forall b \in \Sc_k  \label{eq:H0_td}\\
\mathcal{H}_{1,k}:&\quad
\mathbf{Y}_{b,k} = E_s\,\Jm^{\ell_{b,k}} \mathbf{H}_{b,k} + \mathbf{W}'_{b,k} \;\;\; \forall b \in \Sc_k.  
\label{eq:H1_td}
\end{align}    
\end{subequations}
Then, we use a \ac{glrt} approach \cite{Poor2013Introduction} and first maximize a surrogate likelihood function with respect to $\{\ell_{b,k} : b \in \Sc_k\}$ and finally plug in the estimated delays into the overall likelihood function to decide for hypotheses $\mathcal{H}_{0,k}$ or $\mathcal{H}_{1,k}$. Since the RU received signals are conditionally independent given $a_k$ and the delays, the maximization with respect to the delays can be done separately over $\ell_{b,k} \in \{0, L-1\}$. 
Eventually, the codeword detection reduces to searching, for all shifts $\ell_{b,k}$, the window of samples of length CP containing the larger signal energy. Then, summing over $b \in \Sc_k$ such maximal energies, compare the result with a threshold. By setting the threshold, one can sweep the detector operating curve, trading off false-alarm (FA) and misdetection (MD) probabilities. The mathematical details are omitted due to space limitations. 

\subsection{Frequency-Domain Approach}

We notice that \eqref{eq:fd_matrix_model} is the ``parallel channel'' version (one channel for each subcarrier $\xi$) of the model considered in \cite{Cakmak2025Joint}. Therefore, we shall apply the same multisource AMP algorithm as in \cite{Cakmak2025Joint} to each subcarrier, and pool the outputs together before detection. 
Consider the $N \times BM$ matrix 
\[ \Xm[\xi] = [\Xm_1[\xi], \ldots, \Xm_B[\xi]]. \]
The multisource AMP at iterations $t = 1, 2, \ldots, T$ yields the so-called ``decoupled observation model'' in the form
\begin{equation}
    \Rm^{(t)}[\xi] = \Xm[\xi] + \Psim^{(t)}[\xi],   \label{AMP-output}
\end{equation}
where $\Psim[\xi]$ is a zero-mean Gaussian matrix with i.i.d. components whose variance can be calculated via the AMP \ac{se} for every $t$. The AMP algorithm depends on a denoising function, which is usually obtained as the posterior mean estimation $\mathbb{E}[ \Xm[\xi]  | \Xm[\xi] + \Psim^{(t)}[\xi]]$. This is known as the Bayes-optimal denoiser. However, in our case, the statistics of $\Xm[\xi]$ are not fully known because the rows of $\Xm[\xi]$ are formed by the frequency-domain channels from the active users to the RUs, which depend on many unknown parameters. 
To cope with this lack of information, we notice that the channels are zero-mean vectors (even the LoS components due to the random phase). Hence, we use a mismatched denoiser that treats the channels as Gaussian vectors with a diagonal covariance that depends on the location $u$ and RU index $b$ only, which can be learned from the network topology and the \acp{lsfc}. The results in \cite{Cakmak2025Joint} guarantee that the SE and  \eqref{AMP-output} hold for any Lipschitz denoising function (of course, the noise variance of $\Psim^{(t)}[\xi]$ depends on the chosen denoiser). 

After $T$ AMP iterations, we use the output statistics \eqref{AMP-output}  for $t = T$ and apply the binary 
hypothesis test
\begin{subequations}  \label{composite-hyppo1}
\begin{align}
\mathcal{H}_{0,k}:&\quad
\underline{\rv}^{(T)}_{k} [\xi] = \underline{\psiv}^{(T)}_k[\xi],   \;\;\; \forall \xi \in \Lc  \label{eq:H0_td1}\\
\mathcal{H}_{1,k}:&\quad
\underline{\rv}^{(T)}_{k} [\xi] = \underline{\tilde{\hv}}_k[\xi] + \underline{\psiv}^{(T)}_k[\xi],   \;\;\; \forall \xi \in \Lc  
\label{eq:H1_td1}
\end{align}    
\end{subequations}
to each $k$-th row of $\Rm^{(T)}[\xi]$, where $\underline{\tilde{\hv}}_k[\xi]$ arranged as row vectors is the $1 \times BM$ concatenation of the channels $\tilde{\hv}_{b,k}[\xi]$ and $\Lc$ is the set of pilot subcarriers. For the corresponding likelihood ratio test \cite{Poor2013Introduction}, we notice that the AMP output statistics leading to \eqref{composite-hyppo1} are conditionally independent with respect to the RUs but conditionally dependent over the subcarriers (due to the frequency-correlated channel model).
However, the frequency correlation is a priori unknown for random access users.\footnote{By definition, these are new users joining the system for the first time or after a long idle time.} 
Hence, omitting the superscript $T$ for notational simplicity, we disregard such correlation and consider a mismatched log-likelihood ratio in additive form
\[ \sum_{\xi \in \Lc} \sum_{b \in \Sc_k} \log \big(\Lambda_{b,k}(\underline{\rv}_{b,k} [\xi])\big), \]
where $\Lambda_{b,k}\big(\underline\rv_{b,k}[\xi]\big) \eqdef \frac{p\big(\underline\rv_{b,k}[\xi]\mid a=1\big)}{p\big(\underline\rv_{b,k}[\xi]\mid a=0\big)}$, $\underline{\rv}_{b,k} [\xi]$ is the $b$-th $1 \times M$ section of $\underline{\rv}^{(T)}_{k} [\xi]$. We first derive the non-active and active probabilities for $\underline{\rv}_{b,k}[\xi] = \underline{\xv}_{b,k}[\xi] + \underline{\psiv}_{b,k}[\xi]$ where $\underline{\psiv}_{b,k}[\xi] \sim \Cc\Nc(\underline{\zerov}, \Cm_{b}[\xi])$ with $\Cm_{b}[\xi]$ is obtained by $T$ iterations of the \ac{se} in \cite[Def. 1]{Cakmak2025Joint}. We have $p(\underline{\rv}_{b,k}[\xi]\vert a=0) = \textswab{g}(\underline{\rv}_{b,k}[\xi]; \underline{\zerov}, \Cm_{b}[\xi])$ and since each $k$-th codeword is associated with a location $\Dc_u$ and the RUs with LoS to $\Dc_u$ are known, 
we can exploit the fact that the LoS sections $\underline{\rv}_{b,k} [\xi]$ contain a Rician term (corresponding to the LoS path) with random phase, i.e., $p(\underline{\rv}_{b,k}[\xi]|a=1) = \int_0^{2\pi} p(\underline{\rv}_{b,k}[\xi]|\varphi,a=1)\frac{d\varphi}{2\pi}$, which leads 
\begin{equation}
\begin{aligned}
    p(\underline{\rv}_{b,k}[\xi]|a=\!1)
    &\!=\! \frac{{\rm e}^{-\underline{\rv}_{b,k}[\xi]\Bm_{b}^{-1}[\xi]\underline{\rv}^\Herm_{b,k}[\xi] - \!\underline{\muv}_{b,k}[\xi]\Bm_{b}^{-1}[\xi]\underline{\muv}^\Herm_{b,k}[\xi]}}{\pi^M |\Bm_{b}[\xi]|}\\&\cdot I_0\big(2|z(\underline{\rv}_{b,k}[\xi])|\big),
\end{aligned}
\end{equation}
where $\underline{\muv}_{b,k}[\xi]$ denotes the mean vector, determined solely by the deterministic LoS component, $\Sigmam_{b}[\xi] \in\Cbb^{M\times M}$ is the nominal spatial correlated covariance matrix, \(I_0(\cdot)\) denotes the modified Bessel function of the first kind and order zero, $\Bm_{b}[\xi] \eqdef \big(\Sigmam_{b}[\xi] + \Cm_{b}[\xi]\big)$ and $z(\underline{\rv}_{b,k}[\xi]) \eqdef \underline{\muv}_{b,k}[\xi]\Bm_{b}^{-1}[\xi]\underline{\rv}^\Herm_{b,k}[\xi]$. Then, the likelihood ratio is
\begin{equation}
\begin{aligned}
   \Lambda_{b,k}&(\underline{\rv}_{b,k} [\xi])=\frac{\vert \Cm_{b} [\xi] \vert }{\vert \Bm_{b} [\xi]\vert} {\rm e}^{\underline{\rv}_{b,k} [\xi]\Cm^{-1}_{b} [\xi]\Wm^\Herm_{b} [\xi]\underline{\rv}^\Herm_{b,k} [\xi]}\\&{\rm e}^{- \underline{\muv}_{b,k} [\xi]\Bm_{b}^{-1} [\xi]\underline{\muv}^\Herm_{b,k} [\xi]}I_0\big(2|z(\underline{\rv}_{b,k} [\xi])|\big)
\label{eq:ad_fd}
\end{aligned}
\end{equation}
where $\Wm_{b} [\xi]\eqdef \Bm_{b}^{-1} [\xi]\Sigmam_{b} [\xi]$.

\subsection{Active User Position Estimation}
\label{sec:position_estimation}

First, we notice that only the positioning of the active users effectively detected as active is relevant, because FA events do not correspond to any real user, and MD events are active users that are missed by the detector and therefore ignored (since they do not receive an ACK, they will access the \ac{rach} again after some timeout).  
Then, we build on the idea of \cite{Gkiouzepi2024Joint,Gkiouzepi2025Joint}. In particular, we notice that both the time-domain observation of the type \eqref{eq:mf_out_td} and the frequency-domain observation of the type  \eqref{eq:fd_scalar_model} with $a_k = 1$ (active user) have channel parameters (in particular, the AoA and the delay of the LoS component) that depend on the $k$-th active user position. Furthermore, the $k$-th codeword is associated with a specific location $\Dc_u$ for which the RUs in LoS conditions are known. Hence, we can build a positioning likelihood function using the conditional probability density of the active user observation at the RUs in LoS obtained from the MF receiver (for the time-domain scheme) or from $T$ iterations of the multisource AMP (for the frequency-domain scheme), where now the channel vector is Rician with random phase, and we treat the scattered component as zero-mean Gaussian with a ``nominal'' diagonal covariance that can be estimated from the environment itself.
Then, we evaluate the resulting likelihood function at all points of a fine positioning grid of points $\Gc_u$ in $\Dc_u$ and decide the position using a \ac{mmle} approach. 
We here derive the frequency-domain localization scheme and omit the details of the time-domain scheme due to space limits.

We first stack the per-\ac{ru} per-subcarrier observations into the vector $\underline{\rv}_b \in \Cbb^{1 \times ML_f}$ (with an abuse of notation)
\begin{equation}
    \underline{\rv}_b \eqdef
    \Big[
      \underline{\breve{\hv}}_b[0]  + \underline{\psiv}_b[0],\;
      \ldots,\;
      \underline{\breve{\hv}}_b[L_f-1] + \underline{\psiv}_b[L_f-1]
    \Big].
    \label{eq:rv_deffinal}
\end{equation}
Recall $\breve{\Hm}_{b} \in \Cbb^{L_f \times M}$ of \eqref{eq:breveH}, $\vec(\Am\Bm\Cm)=(\Cm^\Tran\bullet\Am)\vec(\Bm)$, the random vector $\underline{\rv}_b$ of \eqref{eq:rv_deffinal} has the the marginal (phase-averaged) \ac{pdf} of 
\begin{equation}
    p(\underline{\rv}_b)=\frac{e^{-\underline{\rv}_b\Km_b^{-1}\underline{\rv}_b^\Herm -\underline{\muv}_{b,0}\Km_b^{-1}\underline{\muv}_{b,0}^\Herm}}{\pi^{ML_f}|\Km_b|}  I_0\!\Big(2\,\big|\underline{\muv}_{b,0}\Km_b^{-1}\underline{\rv}_b^\Herm\big|\Big),
    \label{eq:marginal_pdf_loc}
\end{equation}
where $\underline{\muv}_{b,0}$ is defined as 
\begin{equation}
    \underline{\muv}_{b,0} =\Ibb^\los_b \sqrt{\beta_{b}^\los}    \Big(\Phim_{b,L_f}^{\ell_0}\bv_b(\mu_0)\Big)^\Tran \bullet\,\underline{\av}(\theta_{b}^\los) \in\Cbb^{1\times ML_f},
    \label{eq:mean_vec_loc}
\end{equation}
with $\bullet$ denotes the Kronecker product and the covariance matrix $\Km_b \in \Cbb^{ML_f\times ML_f}$ is
\begin{equation}
    \Km_b = \Km_b^{\nlos} + \Km_b^{\Psi},
    \label{eq:Kb_decomp}
\end{equation}
admits the structure
\begin{equation}
\label{eq:block_cov}
    \Km_b\!=\!
    \begin{bmatrix}
    \Km_b^{(0,0)} & \Km_b^{(0,1)} & \cdots & \Km_b^{(0,L_f-1)} \\
    \Km_b^{(1,0)} & \Km_b^{(1,1)} & \cdots & \Km_b^{(1,L_f-1)} \\
    \vdots & \vdots & \ddots & \vdots \\
    \Km_b^{(L_f-1,0)} & \Km_b^{(L_f-1,1)} & \cdots & \Km_b^{(L_f-1,L_f-1)}
    \end{bmatrix},\!
\end{equation}
where each block $\Km_b^{(l,l')} \in \Cbb^{M \times M}$ represents the spatial covariance between subcarriers $l$ and $l'$. After calculations omitted for brevity, and by applying the mixed-product property of the Kronecker product and $(\Phim_{b,L_f}^{\ell_p})^* = \Phim_{b,L_f}^{-\ell_p}$, we get
\begin{equation}
    \Km_b^{\nlos}\!=\!\sum_{p \in \Pc} \beta_{b,p} \underbrace{\Phim_{b,L_f}^{-\ell_p}\bv_b^*(\mu_p)\bv_b^\Tran(\mu_p)\Phim_{b,L_f}^{\ell_p}}_{L_f\times L_f}\bullet\underbrace{\underline{\av}^\Herm(\theta_{b,p})\underline{\av}(\theta_{b,p})}_{M\times M},
    \label{eq:Kb_nlos}
\end{equation}
and since we apply the multisource \ac{amp} separately to each subcarrier, the output noise covariance is block-diagonal (subcarriers are independent) 
\begin{equation}
    \Km_b^{\Psi}= \blkdiag\Big(\Cm_b[0],\Cm_b[1],\ldots,\Cm_b[L_f-1]\Big).
    \label{eq:Kb_phi}
\end{equation}

Again, we ignore the correlation across subcarriers in \eqref{eq:block_cov}, and the covariance matrix becomes a block-diagonal matrix
\begin{equation}
\label{eq:block_diag_cov}
    \tilde{\Km}_b=\blkdiag\Big(\tilde{\Km}_b[0],\tilde{\Km}_b[1],\ldots,\tilde{\Km}_b[L_f-1]\Big),
\end{equation}
where each block $\tilde{\Km}_b[\xi] = \Sigmam_b[\xi] + \Cm_b[\xi]$ corresponds to the spatial covariance of the subcarrier $\xi$.

We consider the \ac{mmle} of the user position by minimizing the log-likelihood function of the form
\begin{equation}
\hat\xsf = \arg\min_{\xsf\in\Gc_u}\;\sum_{b\in\Sc_u}\log p(\underline\rv_b\mid\xsf),
\label{eq:fd_pos_est}
\end{equation}
where $\xsf$ is a candidate position. It is also clear that the parameters for the mean components in \eqref{eq:mean_vec_loc} can be re-parameterized and expressed directly in terms of $\xsf$ ranging over a discretization grid $\Gc_u$ of $\Dc_u$. Moreover, we write 
\begin{equation}
    p(\underline{\rv}_b|\xsf)=\frac{e^{-\underline{\rv}_b\tilde{\Km}_b^{-1}\underline{\rv}_b^\Herm -\underline{\muv}_{b,0}\tilde{\Km}_b^{-1}\underline{\muv}_{b,0}^\Herm}}{\pi^{ML_f}|\tilde{\Km}_b|}  I_0\!\Big(2\,\big|\underline{\muv}_{b,0}\tilde{\Km}_b^{-1}\underline{\rv}_b^\Herm\big|\Big).
    \label{eq:marginal_pdf_loc_final}
\end{equation}
We adopt a mismatched covariance $\bar{\Km}_b \in \Cbb^{L_f M \times L_f M}$ formed with matrices $\bar{\Km}_b[\xi] = \bar{\Sigmam}_b + \Cm_b[\xi]$ using the multisource \ac{amp} \ac{se} and the offline estimated diagonal matrices for the covariance from the environment itself.

\section{Simulation Results}
\label{sec:sim_results}

We consider the layout in Fig.~\ref{fig:network_topology} where each hexagon has radius $\unit[100]{m}$, and the RUs have $M = 8$ antennas each.  
The global uRA codebook contains $N=2016$ codewords randomly generated with Gaussian i.i.d. entries and energy-normalized, partitioned uniformly into location-specific subcodebooks of size $N/U = 288$.
For the frequency-domain receiver, we use $Q=144$ CP-OFDM pilot symbols over $L_f=16$ active subcarriers corresponding to a total of $16 \times 144 = 2304$ pilot symbols.
We use a CP of $16$ chips, long enough for the maximum delay spread in the system, which is equal to $\unit[720]{nsec}$ with a system bandwidth of $\unit[20]{MHz}$, yielding a total pilot resource (including CP overhead) of $Q\bigl(L_f+{\rm CP}\bigr)=144\cdot 32=4608$ time-domain chips.
For the time-domain scheme, we consider a codebook of ZC sequences of length $L=2297$, which is then transmitted with a CP of the same length to ensure equal time-frequency resource allocations for both schemes. In this case, the subcodes per location have size $(L-1)/(2*U) = 164$ with a reuse factor of two. In general, the frequency domain scheme allows much more flexibility in terms of codebook dimensioning, thus providing better control of the preamble collision probability, but requires generally larger sequence lengths.
We use the pathloss function given by $\mathrm{PL}(d) = \Big(\frac{\csf}{4\pi f_c d} \Big)^2$ where $f_c$ is the carrier frequency, $d$ is the distance, and the pathloss exponent is equal to 2 since we consider only LoS and specular scatterers. For the specular paths, an additional loss is accounted for through $\sigma_s^2$, which models the scatterer cross-section. The operating frequency is $f_c=\unit[3.5]{GHz}$.
A total of $35$ scatterers are distributed uniformly over each location, each with cross-section $\sigma_s^2 = \unit[-5]{dB}$. The truncation radii for the NLoS are all set equal to $\unit[110]{m}$, while the truncation distance of the LoS is $\unit[200.1]{m}$. The \ac{rach} users' uplink transmit power is determined by fixing a reference receiver \ac{snr}  for a user located at a hexagon center and received at its nearest RU. For a given UE transmit power $P_\mathrm{tx}$, we define the
transmit \ac{snr} as
\begin{equation}
  \mathrm{SNR}_\mathrm{tx}
  = \frac{P_\mathrm{tx}}{W N_0}
  = \frac{\mathcal{E}_s}{N_0},
  \label{eq:snr_def}
\end{equation}
where $\mathcal{E}_s$ is the average energy per time-domain chip and $N_0$ is the noise power spectral density with $N_0=\unit[-174]{dBm/Hz}$. For both schemes we set $\mathcal{E}_s=1$, so $N_0=1/\mathrm{SNR}_\mathrm{tx}$.
The transmit SNR is calculated as the received SNR $\SNR$ divided by the path loss for a user at the center of a hexagon from its nearest \ac{ru}. For the time-domain ZC scheme, each ZC sequence satisfies $\|\sv_k\|^2 = L$, so $\mathcal{E}_s=1$ is achieved directly.
For the frequency-domain scheme, requiring the same per-chip energy as the time-domain scheme is achieved following Parseval's identity since $\Fm$ is unitary, i.e., the two schemes are automatically energy-consistent once their underlying time-domain chip energies are matched. In these results, for each Monte Carlo realization, we generate active user positions, the random selection of their codewords (without preamble collisions), the set of LoS/specular paths, and the channel coefficients according to the model of Section \ref{sec:sys_model}. 
The detection performance is evaluated in terms of MD and FA probabilities. The localization performance is quantified by the Euclidean distance $\|\hat{\pv}_k-\pv_k\|$ restricted to the true-positive set.

\begin{figure}[htb]
	\centering 	\includegraphics[width=0.47\textwidth]{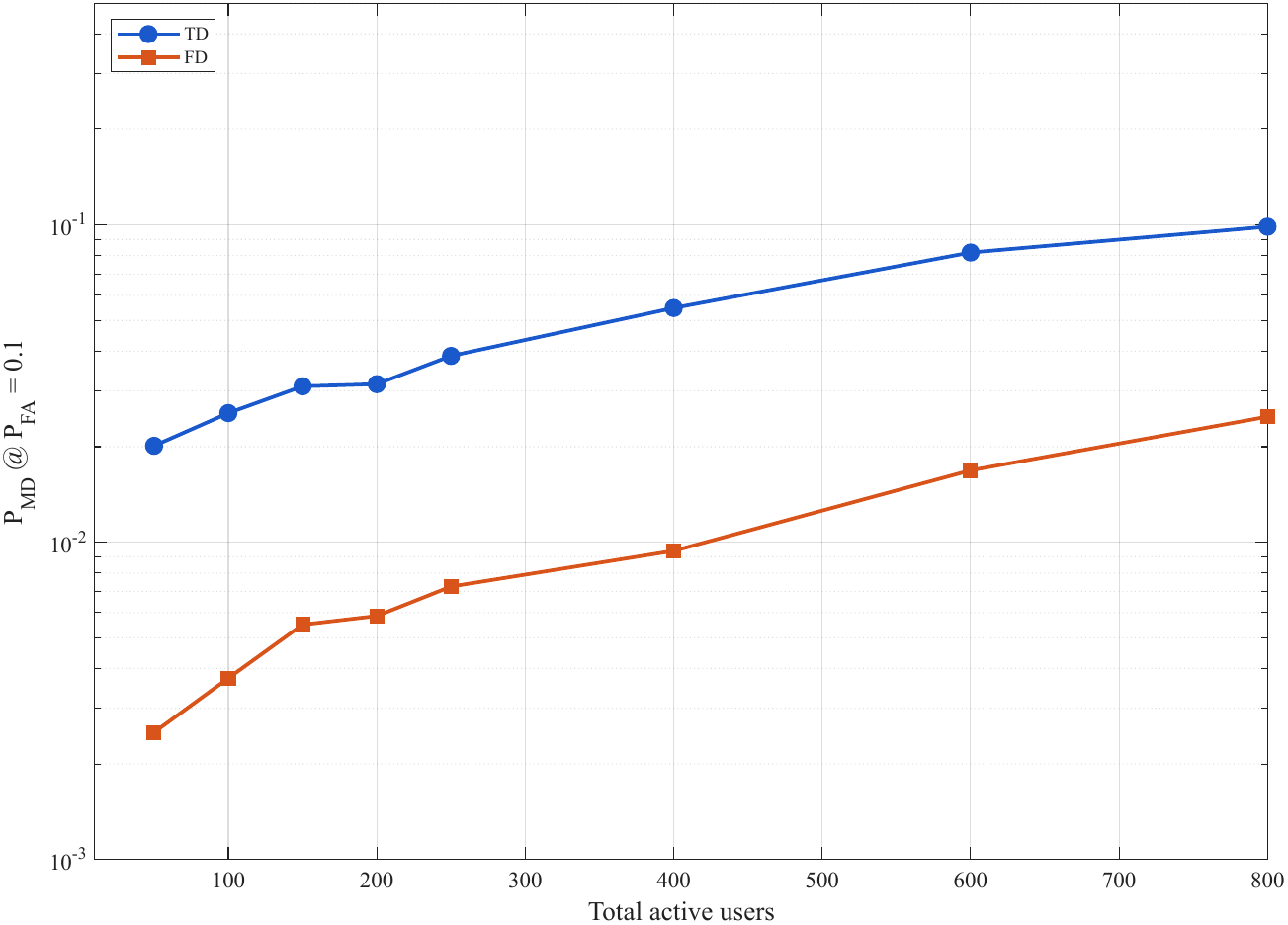}
    
	\caption{MD probability versus the total number of active users in the network at fixed $P_{\rm FA}=10^{-1}$ and at $\SNR = \unit[-30]{dB}$.}\label{fig:pfa_load}	
\end{figure}

Fig.~\ref{fig:pfa_load} shows the MD probability at fixed FA probability $P_{\mathrm{FA}}=10^{-1}$ for increasing random activity (in terms of the total active users in the network). The superiority of the AMP-based frequency domain scheme over the ``legacy'' time-domain scheme is evident.

\begin{figure}[htb]
\centering 	
    \includegraphics[width=0.48\textwidth]{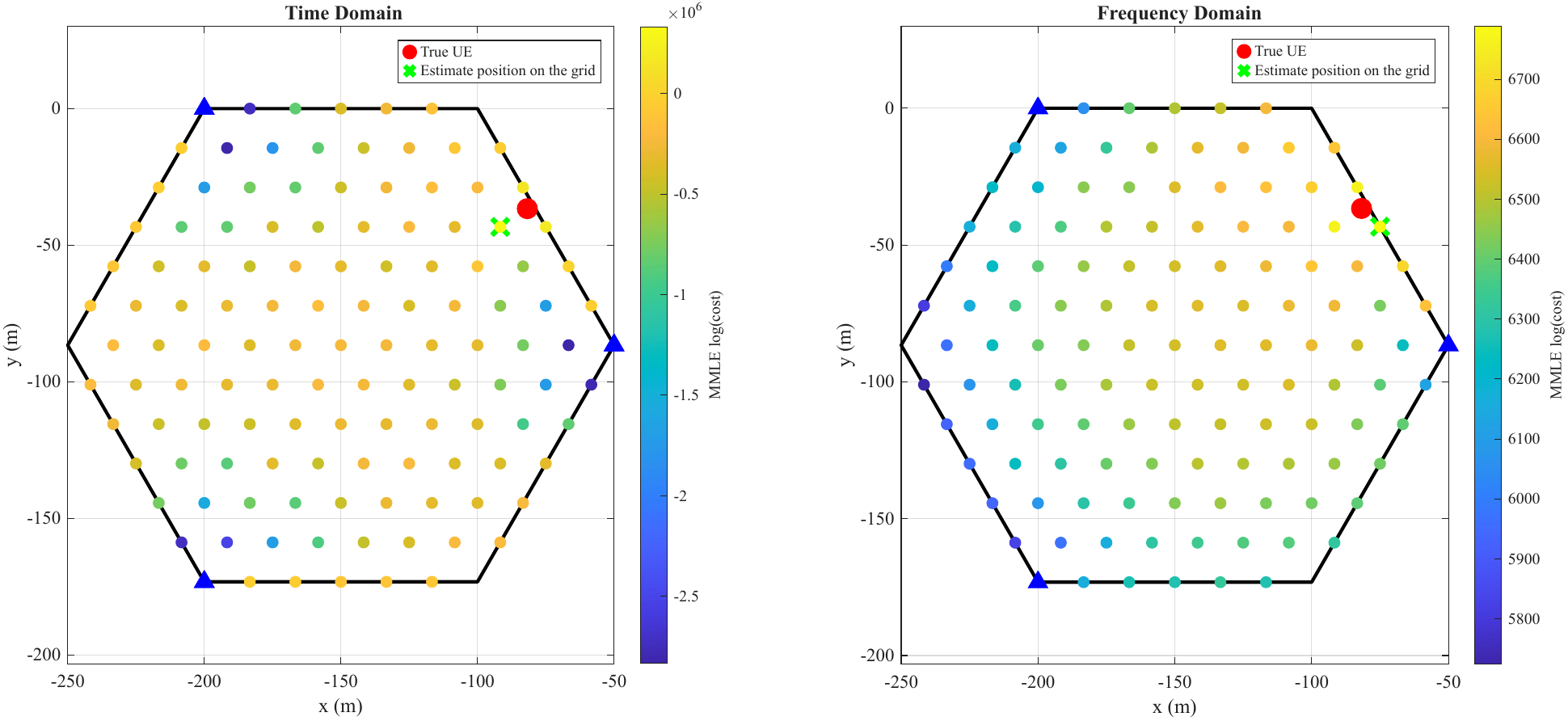}
    \caption{Localization heatmaps over one location $\Dc_u$ for the TD (left) and FD (right) schemes. This also shows the fine grid used for localization.}
    \label{fig:heatmpap}
\end{figure}

\begin{figure}[htb]
	\centering 	\includegraphics[width=0.4\textwidth]{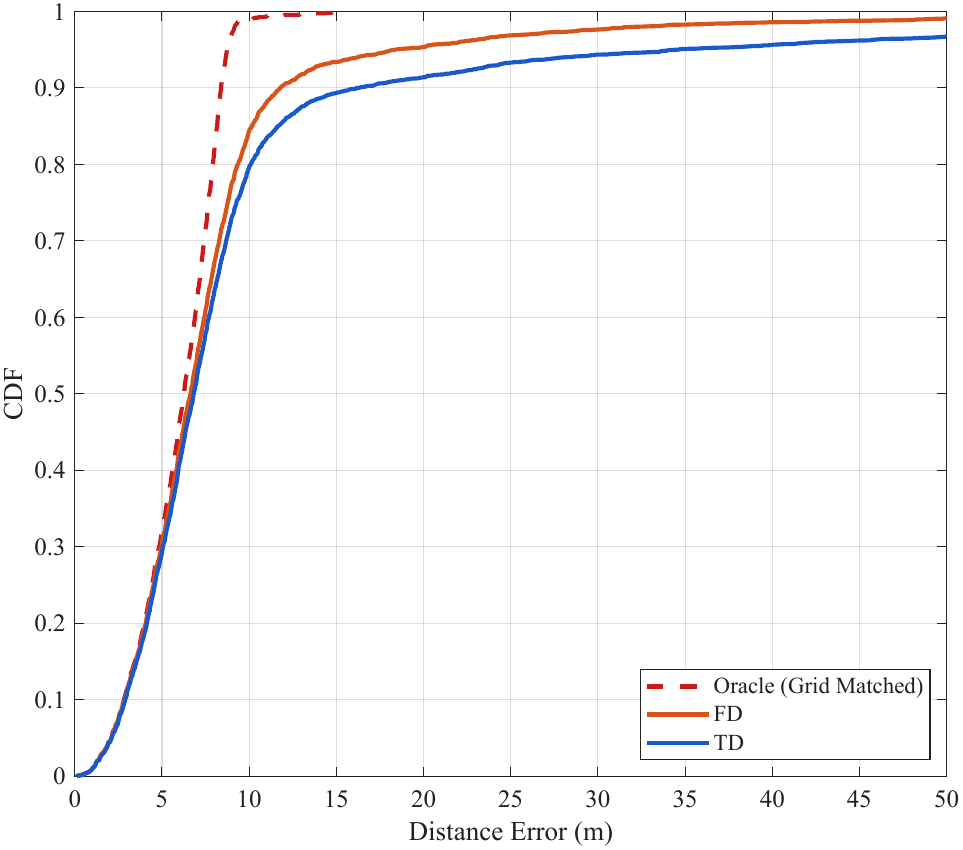}
    \caption{Localization error CDF compared with the oracle benchmark at $\SNR = \unit[10]{dB}$ and 300 random access active users in the network.}
    \label{fig:cdf}
\end{figure}

Fig.~\ref{fig:heatmpap} shows one representative localization snapshot within a single location $\Dc_u$. The estimator is evaluated on a 121-point fine hexagonal grid where the radius of each small grid hexagon is 8.2m. The heat map represents the value of the log-likelihood metric evaluated at each point of the grid. In both cases (for this snapshot), the estimated point is very close to the true point. We report the empirical cumulative distribution function (CDF) of the localization error for both schemes in Fig.~\ref{fig:cdf}. We also compare these with the error CDF of an oracle that always selects the grid point at the minimum distance from the true position, representing the lower bound of the quantization error incurred by the grid. Both the FD and TD schemes exhibit a steep initial rise, indicating a high concentration of small localization errors. However, the FD scheme consistently outperforms the TD scheme, achieving lower localization errors. Specifically, there is a $90\%$ probability that the localization error in the FD scheme is below $\unit[12]{m}$, compared to approximately $\unit[16]{m}$ for the TD scheme.

\section{Conclusion}
\label{sec:Conc}

We studied unsourced random access for the \ac{rach} functionality in cell-free user-centric networks, in a geometrically consistent propagation model including LoS and NLoS propagation paths.   
Two schemes were considered: a ``legacy-like'' time-domain scheme based on a ZC codebook and matched-filter GLRT detection, and a frequency-domain multisource AMP scheme. 
In general, the frequency-domain scheme allows much more flexibility in terms of codebook dimensioning, thus providing better control of the preamble collision probability, but requires generally larger sequence lengths. Moreover, compared to the use of a bank of matched filters and \ac{glrt} test of the time-domain scheme, multisource \ac{amp} jointly processes all codewords, \acp{ru}, antennas, and subcarriers at every iteration, making its computational complexity substantially higher.
For both schemes, the detected active users are localized using an approximate maximum likelihood method, which implicitly makes use of the dependency of delays (TDoA) and AoA from the user position, and naturally operates the sensor fusion of multiple RUs. Future work will address collision-aware models and more refined channel priors for AMP-based inference.

\bibliography{abbrev,references}
\bibliographystyle{IEEEtran}

\end{document}